\documentclass  [journal,onecolumn,tbtags,draftcls,12pt]{IEEEtran}

\usepackage{empheq,amssymb}
\usepackage{braket}
\usepackage{dsfont,mathrsfs}
\usepackage{enumerate}
\usepackage{graphicx}
\usepackage{amsmath}
\usepackage{amsthm}

\usepackage{stfloats}
\newtheorem{problem}    {Problem}
\newtheorem{definition} {Definition}
\newtheorem{claim}      {Claim}
\newtheorem{theorem}    {Theorem}
\newtheorem{lemma}      {Lemma}

\newtheorem{remark} {Remark}

\let \ALPHABET \mathcal

\newcommand\EXP[1]{\mathop{\kern0pt \mathds E}{\Set{#1}}}
\newcommand\PR [1]{\mathop{\kern0pt \Pr}{\Set{#1}}}
\newcommand \RELATION {\mathbf{R}}
\newcommand{\leftexp}[2]%
  {\mathop{}%
   \mathopen{\vphantom{#2}}^{#1}%
   \kern-\scriptspace%
   #2}

\makeatletter
\newcommand\SEQ{\@ifstar\SEQB\SEQA}
\makeatother

\newcommand\SEQA[2][T]{\{#2_t$, $t=1,\dots,#1\}}
\newcommand\SEQB[1]{\{#1_1$, $t=1,\dots\}}

\newcommand{\ind}{\mathds{1}}
\newcommand{\conv}{*}
\newcommand{\norm}[1]{\lVert#1\rVert}

\let\VEC      \mathbf

\title{\LARGE \bf
Optimal Strategies for Communication and Remote Estimation with an Energy Harvesting Sensor}

\author{ A. Nayyar, T. Ba\c{s}ar, D. Teneketzis and V. V. Veeravalli
\thanks{A. Nayyar, T. Ba\c{s}ar and V. Veeravalli are with Coordinated Science Laboratory at the University of Illinois at Urbana-Champaign
        {\tt\small \{anayyar,basar1,vvv\}@illinois.edu}}%
\thanks{D. Teneketzis is with the Department of Electrical Engineering and Computer Science, University of Michigan
        {\tt\small teneket@eecs.umich.edu}}%
}

\begin{document}

\maketitle

\begin{abstract}

We consider a remote estimation problem with an energy harvesting sensor and a remote estimator. The sensor observes the state of a discrete-time  source which may be a finite state Markov chain or a multi-dimensional linear Gaussian system. It harvests energy from its environment (say, for example, through a solar cell) and uses this energy for the purpose of communicating with the estimator.  Due to the randomness of energy available for communication, the sensor may not be able to communicate all the time. The sensor may also want to save its energy for future communications. The estimator relies on messages communicated by the sensor to produce real-time estimates of the source state. We consider the problem of finding a communication scheduling strategy for the sensor and an estimation strategy for the estimator that jointly minimize an expected sum of communication  and distortion costs over a finite  time horizon. Our goal of joint optimization leads to a {decentralized decision-making problem}. By viewing the problem from the estimator's perspective, we obtain a dynamic programming characterization  for the decentralized decision-making problem that involves optimization over functions.  Under some symmetry assumptions on the source statistics and the distortion metric, we  show that an optimal communication strategy is described by easily computable thresholds and that the optimal estimate is a simple function of the most recently received sensor  observation. 

\end{abstract}

\section{Introduction}
Many systems for information collection like sensor networks and environment monitoring networks consist of several network nodes that can observe their environment and communicate with other nodes in the network. Such nodes are typically capable of making \emph{decisions}, that is, they can use the information they have collected from the environment or from other nodes to make decisions about when to make the next observation or when to communicate or how to estimate some state variable of the environment. These decisions are usually made in a \emph{decentralized} way, that is, different nodes make decisions based on different information. Further, such decisions must be made under resource constraints. For example, a wireless node in the network must decide when to communicate under the constraint that it has a limited battery life. In this paper, we study one such  decentralized decision making problem under energy constraints.
\par
We consider a setup where one sensor is observing an environmental process of interest which must be communicated to a remote estimator. The estimator needs to produce estimates of the state of the environmental process in real-time.  We assume that communication from sensor to the estimator is energy consuming. The sensor is assumed to be harvesting energy from the environment (for example, by using a solar cell). Thus, the amount of energy available at the sensor is a random process. Given the limited and random availability of energy, the sensor has to decide when to communicate with the estimator. Given that the sensor may not communicate at all times,  the estimator has to decide how to estimate the state of the environmental process. Our goal is to study the effects of randomness of energy supply on the nature of optimal communication scheduling and estimation strategies. 
\par
Communication problems with energy harvesting transmitters have been studied recently (see \cite{Ulukus1}, \cite{Ulukus2} and references therein). In these problems the goal is to vary the transmission rate/power according to the energy availability in order to maximize throughput and/or to minimize transmission time. In our problem, on the other hand, the goal is to jointly optimize the communication scheduling and the estimation strategies in order to minimize an accumulated communication and estimation cost. 
Problems of communication scheduling and estimation  with a fixed bound on the number of transmissions, independent identically distributed (i.i.d.) sources and without energy harvesting have been studied in \cite{ImerBasar1} and \cite{ImerBasar2}, where scheduling strategies are restricted to be threshold based. A continuous time version of the problem with Markov state process and a fixed number of transmissions is studied in \cite{Rabi_Baras}. In \cite{hespanha}, the authors find an optimal communication schedule assuming a Kalman-like estimator. Remote estimation of a scalar linear Gaussian source with communication costs has been studied in \cite{Martins:2011}, where the authors proved  that a threshold based communication schedule and a Kalman-like estimator are jointly optimal.  Our analytical approach borrows extensively from the arguments in \cite{Martins:2011}  and \cite{Hajek:2008}. The latter considered  a problem of paging and registration in a cellular network which can be viewed as a remote estimation problem. 
\par
Problems where the estimator decides when to query a sensor or which sensor to query have been studied in \cite{arapostathis}, \cite{Sinopoli}, \cite{Gupta}, \cite{geromel}, \cite{Athans}. In these problems, the decision making is centralized. Our problem differs from these setups because the decision to communicate is made by the sensor that has more information than the estimator and this leads to a decentralized decision making problem.
\par
In order to appreciate the difficulty of joint optimization of communication and estimation strategies, it is important to recognize the role of signaling in estimation. When the sensor makes a decision on whether to communicate or not based on its observations of the source, then a decision of \emph{not to communicate conveys information to the estimator}. 
 For example, if the estimator knows that the sensor always communicates if the source state is outside an interval $[a,b]$, then not receiving any communication from the sensor reveals to the estimator that the state must have been inside the interval $[a,b]$. Thus, even if the source is Markov, the estimator's estimate may not simply be a function of the most recently received source state since each successive ``no communication'' has conveyed some information. It is this aspect of the problem that makes derivation of jointly optimal  communication and estimation strategies a difficult problem.

\subsection{Notation}
Random variables are denoted by upper case letters ($X, \Gamma, \Pi, \Theta$), their realizations
by the corresponding lower case letters ($x,\gamma,\pi,\theta$). The notation $X_{a:b}$ denotes 
the vector $(X_a, X_{a+1}, \dots, X_b)$. 
 Bold capital letters $\VEC X$ represent random vectors, while bold small letters $\VEC x$ represent their realizations. $\mathds{P}(\cdot)$ is the
probability of an event, $\mathds{E}(\cdot)$ is the expectation of a random
variable.  $\mathds{1}_A(\cdot)$ is the indicator
function of a set $A$. $\mathbb{Z}$ denotes the set of integers, $\mathbb{Z}_+$ denotes the set of positive integers, $\mathbb{R}$ is the set of real numbers and $\mathbb{R}^n$ is the $n$- dimensional Euclidean space. $\VEC I$ denotes the identity matrix.  
For two random variables (or random vectors) $X$ and $Y$ taking values in
$\ALPHABET X$ and $\ALPHABET Y$, $\mathds{P}(X=x|Y)$ denotes the  conditional
probability of the event $\{X=x\}$ given $Y$ and $\mathds{P}(X|Y)$ denotes the
conditional PMF (probability mass function) or conditional probability density of $X$ given $Y$. These conditional probabilities are 
random variables whose realizations depend on realizations of $Y$.


\subsection{Organization}
In Section \ref{sec:PF}, we formulate our problem for a discrete source. We present a dynamic program for our problem in Section \ref{sec:prelim}. This dynamic program involves optimization over a function space. In Section \ref{sec:results}, we find optimal strategies under some symmetry assumptions on the source  and the distortion function. We consider the multi-dimensional Gaussian source in Section \ref{sec:multi_d}. We present some important special cases in Section \ref{sec:special}. We conclude in Section \ref{sec:conc}. We provide some auxiliary results and proofs of key lemmas in Appendices A to E. This work is an extended version of \cite{Nayyar-CDC12}.
\section{Problem Formulation} \label{sec:PF}
\subsection{The System Model}
Consider a remote estimation problem with a sensor and a remote estimator. The sensor observes a discrete-time Markov process $X_t$, $t=1,2,\ldots$. The state space of this source process is a finite interval $\mathcal{X}$ of the set of integers $\mathbb{Z}$.  The estimator relies on  messages communicated by the sensor to produce its estimates of the process $X_t$. The sensor harvests energy from its environment (say, for example, through a solar cell) and uses this energy for  communicating with the estimator. Let $E_t $ be the  energy level at the sensor at the beginning of time $t$. We assume that the energy level is discrete and takes values in the set $\mathcal{E} = \{0,1,\ldots,B\}$, where $B \in \mathbb{Z}_+$.  In the time-period $t$, the sensor harvests a random amount $N_t$ of energy from its environment, where $N_t$ is a random variable taking values in the set $\mathcal{N} \subset \mathbb{Z}_+$. The sequence  $N_t$, $t =1,2,\ldots,$ is an i.i.d. process which is independent of the source process $X_t$, $t=1,2,\ldots$.

We assume that a successful transmission from the sensor to the estimator consumes $1$ unit of energy. Also, we assume that the sensor consumes no energy if it just observes the source but does not transmit anything to the estimator. At the beginning of the time period $t$, the sensor makes a decision about whether to transmit its current observation and its current energy level to the estimator or not. We denote by $U_t \in \{0,1\}$ the sensor's decision at time $t$, where $U_t=0$ means no transmission and $U_t=1$ means a decision to transmit. Since the sensor needs at least $1$ unit of energy for transmission, we have the constraint that $U_t \leq E_t$. Thus, if $E_t = 0$, then $U_t$ is necessarily $0$. The energy level of the sensor at the beginning of the next time step can be written as
\begin{equation}\label{eq:energy}
E_{t+1} = \min\{E_t +N_t -U_t, B\},
\end{equation}
where $B$ is the maximum number of units of energy that the sensor can store.
The estimator receives a message $Y_t$ from the sensor where
\begin{align}
Y_t = \left \{ \begin{array}{ll}
               (X_t,E_t) & \mbox{if $U_t =1$} \\
               \epsilon & \mbox{ if $U_t=0$ },
               \end{array}
               \right.
\end{align} 
where $\epsilon$ denotes that no message was transmitted.
The estimator produces an estimate $\hat{X}_t$ at time $t$ depending on the sequence of messages it received so far. The system operates for a finite time horizon $T$.
\subsection{Decision Strategies}
The sensor's decision at time $t$ is chosen as a function of its observation history, the history of energy levels and the sequence of past messages. We allow randomized strategies for the sensor (see Remark 1). Thus, at time $t$, the sensor makes the decision $U_t=1$ with probability $p_t$ where 
\begin{equation}
p_t= f_t(X_{1:t},E_{1:t},Y_{1:t-1})
\end{equation}  
The constraint $U_t \leq E_t$ implies that we have the constraint that $p_t=0$ if $E_t =0$.
The function $f_t$ is called the decision rule of the sensor at time $t$ and the collection of functions $ \mathbf{f} = \{f_1,f_2,\ldots,f_T\}$ is called the decision strategy of the sensor.
\par
The estimator produces its estimate as a function of the messages,
\begin{equation}
\hat{X}_t= g_t(Y_{1:t})
\end{equation}  
The function $g_t$ is called the decision rule of the estimator at time $t$ and the collection of functions $ \mathbf{g} = \{g_1,g_2,\ldots,g_T\}$ is called the decision strategy of the estimator.

\subsection{The Optimization Problem}
We have the following optimization problem.

\begin{problem}
  \label{prob:main}
  For the model described above, given the statistics of the Markov source and the initial energy level $E_1$, the statistics of amounts of energy harvested at each time, the sensor's energy storage limit $B$ and the time horizon $T$,  find  decision strategies $\VEC f, \VEC g$ for the sensor and the estimator, respectively, that minimize the following expected cost:
  \begin{equation}\label{eq:cost}
  J(\VEC f, \VEC g) = \mathds{E} \{\sum_{t=1}^T c U_t + \rho(X_t,\hat X_t) \},
  \end{equation}
 where $c \geq 0$ is a communication cost and $\rho: \mathcal{X} \times \mathcal{X} \mapsto \mathbb{R}$ is a distortion function. 
\end{problem}

\begin{remark}
It can be argued that in the above problem, sensor strategies can be assumed to be deterministic (instead of randomized) without compromising  optimality.  However, our argument  for characterizing optimal strategies makes use of the possibility of randomizations by the sensor and therefore we allow for randomized strategies for the sensor. 
\end{remark}

\textbf{Discussion of Our Approach:} Our approach for Problem \ref{prob:main} makes extensive use of majorization theory based arguments used in \cite{Hajek:2008} and \cite{Martins:2011}. As in \cite{Hajek:2008}, we first construct a dynamic program for Problem \ref{prob:main} by reformulating the problem from the estimator's perspective. This dynamic program involves minimization over a function space. Unlike the approach in \cite{Hajek:2008}, we use majorization theory to argue that the value functions of this dynamic program, under some symmetry conditions, have a special property that is similar to (but not the same as) Schur-concavity \cite{Marshall}. We then use this property to characterize the solution of the dynamic program. This characterization then enables us to find  optimal strategies. In Section \ref{sec:multi_d}, we consider the problem with a multi-dimensional Gaussian source. We extend our approach for the discrete case to this problem  and, under a suitable symmetry condition, we provide optimal strategies for this case as well.  While the result in \cite{Martins:2011} is only for scalar Gaussian source without energy harvesting, our approach addresses multi-dimensional source and energy harvesting. Finally, in Section \ref{sec:special}, we mention a few special cases which include the important  remote estimation problems where the sensor can afford only a fixed number of transmissions or where the sensor only has a communication cost but no constraint on the number of transmissions.
\section{Preliminary Results}\label{sec:prelim}
\begin{lemma}\label{lemma:str}
 There is no loss of performance if the sensor is restricted to decision strategies  of the form:
 \begin{equation}\label{eq:lemma_1}
 p_t = f_t(X_t,E_t, Y_{1:t-1})
 \end{equation}
 \end{lemma}
 \begin{proof} Fix the estimator's strategy $\VEC g$ to any arbitrary choice. We will argue that, for the fixed choice of $\VEC g$, there is an optimal sensor strategy of the form in the lemma. To do so, we can show that with a fixed $\VEC g$ the sensor's optimization problem is a Markov decision problem with $X_t,E_t,Y_{1:t-1}$ as the state of the Markov process. It is straightforward to establish that conditioned on $X_t,E_t,Y_{1:t-1}$ and $p_t$, the next state $(X_{t+1},E_{t+1},Y_{1:t})$ is independent of past source states and energy levels and past choices of transmission probabilities. Further, the expected cost at time $t$  is a  function of the state and $p_t$. Thus, the sensor's optimization problem is a Markov decision problem with $X_t,E_t,Y_{1:t-1}$ as the state of the Markov process.   Therefore, using standard results from Markov decision theory \cite{KumarVaraiya:1986}, it follows that an optimal sensor strategy is of the form in the lemma. Since the structure of the sensor's optimal strategy is true for an arbitrary choice of $\VEC g$, it is also true for the globally optimal choice of $\VEC g$. This establishes the lemma. 
 \end{proof}  
  
 In the following analysis, we will  consider only sensor's strategies of the form in Lemma~\ref{lemma:str}.
 Thus, at the beginning of a time instant $t$ (before the transmission at time $t$ happens), the sensor only needs to know $X_t$, $E_t$ and $Y_{1:t-1}$, whereas the estimator knows $Y_{1:t-1}$. Problem~\ref{prob:main} - even with the sensor's strategy restricted to the form in Lemma \ref{lemma:str}- is a decision-problem with \emph{non-classical information structure} \cite{Ho:1980}. One approach for addressing such problems is to view them from the perspective of a decision maker who knows only the common information among the decision makers \cite{mythesis}. In Problem \ref{prob:main}, at the beginning of time $t$, the information at the sensor is ($X_t,E_t,Y_{1:t-1}$), while the information at the estimator is $Y_{1:t-1}$. Thus, the estimator knows the common information ($Y_{1:t-1}$) between the sensor and the estimator. We will now formulate a decision problem from the estimator's point of view and show that it is equivalent to Problem~\ref{prob:main}.
\subsection{An Equivalent Problem} \label{sec:equiv prob}
   We formulate a new problem in this section. Consider the model of Section \ref{sec:PF}. At the end of time $t-1$, using the information $Y_{1:t-1}$, the estimator decides an estimate
 \begin{align}
 \hat{X}_{t-1} &= g_t(Y_{1:t-1}) \notag
 \end{align}
 In addition, at the beginning of time $t$, the \emph{estimator} decides a \emph{function} $\Gamma_t: \mathcal{X} \times \mathcal{E} \mapsto [0,1]$, using the information $Y_{1:t-1}$. That is,
 \begin{align}
 \Gamma_t &= \ell_t(Y_{1:t-1}).
 \end{align}
 Then, at time $t$, the sensor evaluates its transmission probability as $p_t= \Gamma_t(X_t,E_t)$. We refer to $\Gamma_t$ as the \emph{prescription} to the sensor. The sensor simply uses the prescription to evaluate its transmission probability. The estimator can select a prescription from the set $\mathcal{G}$, which is the set of all functions $\gamma$ from $\mathcal{X} \times \mathcal{E}$ to $[0,1]$ such that $\gamma(x,0) =0, \forall x \in \mathcal{X}$. It is clear that any prescription in the set $\mathcal{G}$ satisfies the energy constraint of the sensor, that is, it will result in $p_t=0$ if $E_t=0$. We call $\boldsymbol {\ell} := \ell_1,\ell_2,\ldots,\ell_T$ the \emph{prescription strategy} of the estimator. Thus, in this formulation, the estimator is the only decision maker. This idea of viewing the communication and estimation problem only from the estimator's perspective has been used in \cite{Walrand83}, \cite{Hajek:2008}. A more general treatment of this approach of viewing problems with multiple decision makers from the viewpoint of an agent who knows only the common information can be found in \cite{mythesis}. We can now formulate the following optimization problem for the estimator.
 
   \begin{problem}\label{prob:2}
    For the model described above, given the statistics of the Markov source and the initial energy level $E_1$, the statistics of amounts of energy harvested at each time, the sensor's energy storage limit $B$ and the time horizon $T$,  find an estimation strategy $\VEC g$, and a prescription strategy $\boldsymbol {\ell}$ for the the estimator that minimizes the following expected cost:
  \begin{equation}\label{eq:cost2}
  \hat J(\boldsymbol \ell, \VEC g) = \mathds{E} \{\sum_{t=1}^T c U_t + \rho(X_t,\hat X_t) \},
  \end{equation}
\end{problem} 
\par
Problems~\ref{prob:main} and \ref{prob:2} are equivalent in the following sense: Consider any choice of strategies $\VEC f, \VEC g$ in Problem~\ref{prob:main}, and define a prescription strategy in Problem~\ref{prob:2} as
\[ \ell_t(Y_{1:t-1}) = f_t(\cdot,\cdot,Y_{1:t-1}) \]
Then, the strategies $\boldsymbol \ell, \VEC g$ achieve the same value of the total expected cost in Problem~\ref{prob:2} as the strategies $\VEC f ,\VEC g$ in Problem~\ref{prob:main}. Conversely, for any choice of strategies $\boldsymbol \ell, \VEC g$ in Problem~\ref{prob:2}, define a sensor's strategy in Problem~\ref{prob:main} as
\[ f_t(\cdot,\cdot,Y_{1:t-1}) = \ell_t(Y_{1:t-1}) \]
Then, the strategies $\VEC f, \VEC g$ achieve the same value of the total expected cost in Problem~\ref{prob:main} as the strategies $\boldsymbol \ell ,\VEC g$ in Problem~\ref{prob:2}.
\par
 Because of the above equivalence, we will now focus on the estimator's problem of selecting its optimal estimate and the optimal prescriptions (Problem~\ref{prob:2}). We will then use the solution of Problem~\ref{prob:2} to find optimal strategies in Problem~\ref{prob:main}. 
 
Recall that $E_t$ is the sensor's energy level at the beginning of time $t$. For ease of exposition, we define a post-transmission energy level at time $t$ as $E'_t = E_t-U_t$. The estimator's optimization problem can now be described as a partially observable Markov decision problem  (POMDP) as follows:
\begin{enumerate}
\item State processes:   $  (X_t,E_t)$ is the pre-transmission state;  $ (X_t,E'_t)$ is the post-transmission state. 
\item Action processes:  $  \Gamma_t$ is the pre-transmission action;  $ \hat X_t$ is the post-transmission action.

\item Controlled  Markovian Evolution of States: The state evolves from $(X_t,E_t)$ to $(X_t,E'_t)$ depending on the realizations of $X_t,E_t$ and the choice of pre-transmission action $\Gamma_t$. The post-transmission state is $(X_t,E_t-1)$ with probability $\Gamma_t(X_t,E_t)$ and $(X_t,E_t)$ with probability $1-\Gamma_t(X_t,E_t)$. The state then evolves in a Markovian manner from $(X_t,E'_t)$ to $(X_{t+1},E_{t+1})$ according to known statistics that depend on the transition probabilities of the Markov source and the statistics of the energy harvested at each time.

\item Observation Process: $Y_t$. The observation is a function of the pre-transmission state and the pre-transmission action. The observation is $(X_t,E_t)$ with probability $\Gamma_t(X_t,E_t)$ and $\epsilon$ with probability $1-\Gamma_t(X_t,E_t)$.

\item Instantaneous Costs: The communication cost at each time is a function of the pre-transmission state and the pre-transmission action. The communication cost is $c$ with probability $\Gamma(X_t,E_t)$ and $0$ with probability $1-\Gamma(X_t,E_t)$. The distortion cost at each time step, $\rho(X_t,\hat{X}_t)$ is a function of the post-transmission state and the post-transmission action.
\end{enumerate}
\par
The above equivalence with POMDPs suggests that the estimator's posterior beliefs on the states are its information states \cite{KumarVaraiya:1986}.
We, therefore, define the following probability mass functions (PMFs):
\begin{definition}
\begin{enumerate}
\item  We define the pre-transmission belief at time $t$ as $\Pi_t := \mathds{P}(X_t,E_t|Y_{1:t-1})$. Thus, for $(x,e) \in \mathcal{X} \times \mathcal{E}$, we have \[\Pi_t(x,e)=\mathds{P}(X_t=x,E_t=e|Y_{1:t-1}).\]
\item  We define the post-transmission belief at time $t$ as $\Theta_t := \mathds{P}(X_t,E'_t|Y_{1:t})$. Thus, for $(x,e) \in \mathcal{X} \times \mathcal{E}$, we have \[\Theta_t(x,e)=\mathds{P}(X_t=x,E'_t=e|Y_{1:t}).\] 
\end{enumerate}
\end{definition}
The following lemma describes the evolution of the beliefs $\Pi_t$ and $\Theta_t$ in time.
\begin{lemma}\label{lemma:update}
The estimator's beliefs evolve according to the following fixed transformations:
 \begin{enumerate}
\item $ \Pi_{t+1}(x,e) =\sum_{\substack{x' \in \mathcal{X}, \\e' \in \mathcal{E}}} [\mathds{P}(X_{t+1} =x|X_t = x') \mathds{P}(E_{t+1} =e|E'_t = e') \Theta_t(x',e')]. $\\
 We denote this transformation by 
$\Pi_{t+1} = Q^1_{t+1}(\Theta_t)$.
\item 
\begin{align}
\Theta_t(x,e) =  \left \{ \begin{array}{ll}
               \delta_{\{x',e'-1\}}~~~ ~~~~~~~~~~~~ \mbox{if $Y_t =(x',e')$} \\
                \frac{(1-\Gamma_t(x,e))\Pi_t(x,e)}{\sum_{x',e'}(1-\Gamma_t(x',e'))\Pi_t(x',e')}  \mbox{ if $Y_t=\epsilon$ }
               \end{array},
               \right.
\end{align} 
where $\delta_{\{x',e'-1\}}$ is a degenerate distribution  at $(x',e'-1)$.
 We denote this transformation by  $\Theta_t = Q^2_t(\Pi_t,\Gamma_t,Y_t)$. 
\end{enumerate}
\end{lemma}

We can now describe the optimal strategies for the estimator.
%
\begin{theorem}\label{thm:val_functions}
Let $\pi, \theta$ be any PMF defined on $\mathcal{X} \times \mathcal{E}$. Define recursively the following functions:
\begin{equation*}
W_{T+1}(\pi) := 0
\end{equation*}
\begin{align}\label{eq:dpeq1}
&V_t(\theta) := \min_{a \in \mathcal{X}} \mathds{E}[\rho(X_t, a) + W_{t+1}(\Pi_{t+1})|\Theta_t=\theta] 
\end{align}
where $\Pi_{t+1} = Q^1_{t+1}(\Theta_t)$ (see Lemma \ref{lemma:update}), and 
\begin{align} \label{eq:dpeq2}
&W_{t}(\pi) := \min_{\tilde\gamma \in \mathcal{G}} \mathds{E}[c\ind_{\{U_t=1\}}+V_{t}(\Theta_{t})|\Pi_t=\pi,\Gamma_t = \tilde\gamma] 
\end{align}
where $\Theta_{t} = Q^2_{t}(\Pi_t,\Gamma_t,Y_t)$ (see Lemma \ref{lemma:update}).
\par 
For each realization  of the post-transmission belief at time $t$, the minimizer in \eqref{eq:dpeq1} exists and gives the optimal estimate  at time $t$; for each realization of the pre-transmission belief, the minimizer in \eqref{eq:dpeq2} exists  and gives the optimal prescription at time $t$. 
\end{theorem}
\begin{proof}
The  minimizer in \eqref{eq:dpeq1} exists because $\mathcal{X}$ is finite; the minimizer in \eqref{eq:dpeq2} exists because the conditional expectation on the right hand side of \eqref{eq:dpeq2} is a continuous function of $\tilde\gamma$ and $\mathcal{G}$ is a compact set. The optimality of the minimizers follow from standard dynamic programming arguments for POMDPs.
\end{proof}

 The result of Theorem~\ref{thm:val_functions} implies that we can solve the estimator's problem of finding optimal estimates and prescriptions by finding the minimizers in equations \eqref{eq:dpeq1} and \eqref{eq:dpeq2} in a backward inductive manner. Recall that  the minimization in equation \eqref{eq:dpeq2} is over the space of functions in $\mathcal{G}$. This is a difficult minimization problem. In the next section, we  consider a special class of sources and distortion functions that satisfy certain symmetry assumptions. We do not solve the dynamic program but instead use it to \emph{characterize optimal strategies of the sensor and the estimator.} Such a characterization provides us with an alternative way of finding optimal strategies of the sensor and the estimator.
\section{Characterizing Optimal Strategies}\label{sec:results}
\subsection{Definitions}
\begin{definition}
 A probability distribution $\mu$ on $\mathbb{Z}$ is said to be \emph{almost symmetric and unimodal (a.s.u.)} about a point $a \in \mathbb{Z}$, if for any $k = 0,1,2,\ldots,$
 \begin{equation}
 \mu(a+k) \geq \mu(a-k) \geq \mu(a+k+1)
 \end{equation}
 If a distribution $\mu$ is a.s.u. about $0$ and $\mu(x) = \mu(-x)$, then $\mu$ is said to be a.s.u. and even. Similar definitions hold if $\mu$ is a sequence, that is, $\mu: \mathbb{Z} \mapsto \mathds{R}$.
\end{definition}

\begin{definition} \label{def:neatsource}
We call a source \emph{neat} if the following assumptions hold:
\begin{enumerate}
\item The a priori probability of the initial state of the source $\mathds{P}(X_1)$ is a.s.u. and even and  has finite support. 
\item The time evolution of the source is given as:
 \begin{equation} \label{eq:evolution}
  X_{t+1} = X_t + Z_t 
 \end{equation}
 where $Z_t, t=1,2,\ldots, T-1$ are i.i.d random variables with a finite support, a.s.u. and even distribution $\mu$.
 \end{enumerate}
\end{definition}
\begin{remark}
Note that the finite support of the distributions of $X_1$ and $Z_t$ and the finiteness of the time horizon $T$ imply that the state of a neat source always lies within a finite interval in $\mathbb{Z}$. This finite interval is the state space $\mathcal{X}$.
\end{remark}

 We borrow the following notation and definition from the theory of majorization.
\begin{definition}
Given $\mu \in \mathbb{R}^{n}$, let $\mu_{\downarrow} = (\mu_{[1]},\mu_{[2]},\ldots,\mu_{[n]})$ denote the non-increasing rearrangement of $\mu$ with $\mu_{[1]}\geq \mu_{[2]} \geq\ldots \geq \mu_{[n]}$. Given two vectors $\mu$ and $\nu$ from $\mathbb{R}^{n}$, we say that $\nu$ majorizes $\mu$, denoted by, $\mu \prec \nu$, if the following conditions hold:
\begin{align*}
&\displaystyle\sum_{i=1}^{k} \mu_{[i]} \leq \displaystyle\sum_{i=1}^{k} \nu_{[i]}, \mbox{~~~~~~~~~for $1 \leq k \leq n-1$} \\
&\displaystyle\sum_{i=1}^{n} \mu_{[i]} = \displaystyle\sum_{i=1}^{n} \nu_{[i]}
\end{align*}
\end{definition}

We now define a relation $\RELATION$ among possible information states and a property $\RELATION$ of real-valued functions of information states.
\begin{definition} [Binary Relation $\RELATION$]
Let $\theta$ and $\tilde\theta$ be two distributions on $\mathcal{X}\times \mathcal{E}$. We say $\theta\RELATION\tilde\theta$ iff:
\begin{enumerate}
\item [(i)] For each $e \in \mathcal{E}$, 
$\theta(\cdot,e) \prec \tilde\theta(\cdot,e)$
\item [(ii)] For all $e \in \mathcal{E}$, $\tilde\theta(\cdot,e)$ is a.s.u. about the same point $ x \in \mathcal{X}$. 

\end{enumerate}
\end{definition}

\begin{definition}[Property $\RELATION$]
 Let $V$ be a function that maps distributions on $\mathcal{X}\times\mathcal{E}$ to the set of real numbers $\mathbb{R}$. We say that $V$ satisfies Property $\RELATION$ iff for any two distributions $\theta$ and $\tilde\theta$,
 \[\theta \RELATION \tilde\theta \implies V(\theta) \geq V(\tilde\theta)\]
 
\end{definition}

\subsection{Analysis}

In this section, we will consider Problem \ref{prob:main} under the assumptions that:
\begin{enumerate}
\item [(A1)]The source is neat (see Definition \ref{def:neatsource}), and
\item [(A2)] The distortion function $\rho(x,a)$ is either $\rho(x,a) = \ind_{\{x\neq a\}}$ or $\rho(x,a) = |x-a|^k$, for some $k>0$.  
\end{enumerate}
Throughout the following analysis, we will assume that Assumptions A1 and A2 hold.
\begin{lemma} \label{lemma:est result}
Let $\theta$ be a distribution on $\mathcal{X}\times\mathcal{E}$ such that for all $e \in \mathcal{E}$, $\theta(\cdot,e)$ is a.s.u. about the same point $x' \in \mathcal{X}$. Then, the minimum in \eqref{eq:dpeq1} is achieved at $x'$. 
\end{lemma}
\begin{proof}
Using Lemma \ref{lemma:update}, the expression in \eqref{eq:dpeq1} can be written as
\[ V_t(\theta) := W_{t+1}(Q^1_{t+1}(\theta)) + \min_{a \in \mathcal{X}} \mathds{E}[\rho(X_t, a) |\Theta_t=\theta]\]
Thus, the minimum is achieved at the point that minimizes the expected distortion function $\rho(X_t,a)$ given that $X_t$ has the distribution $\theta$. The a.s.u. assumption of all  $\theta(\cdot,e)$  about $x'$ and the nature of distortion functions given in Assumption A2 implies that $x'$ is the minimizer. 
\end{proof}
 
We now want to characterize the minimizing $\tilde\gamma$ in \eqref{eq:dpeq2}. Towards that end, we start with the following claim.
\begin{claim} \label{claim:one}
The value functions $W_t$, $t=1,2,\ldots T+1$, and  $V_t, t=1,2,\ldots,T$, satisfy Property $\RELATION$.

\end{claim}
\begin{proof}
See Appendix \ref{sec:claimproof}.
\end{proof}
Recall that \eqref{eq:dpeq2} in  the dynamic program for the estimator defines $W_t$ as
\begin{align} 
W_{t}(\pi) := \min_{\tilde\gamma} \mathds{E}[c\ind_{\{U_t=1\}}+V_{t}(\Theta_{t})|\Pi_t=\pi,\gamma_t = \tilde\gamma] 
\end{align}
The following lemma is a consequence of Claim~\ref{claim:one}.
\begin{lemma}\label{lemma:thresh1}
Let $\pi$ be a distribution on $\mathcal{X}\times\mathcal{E}$ such that $\pi(\cdot,e)$ is a.s.u. about the same point $a \in \mathcal{X}$ for all $e \in \mathcal{E}$. Then, the minimum in the definition of $W_t(\pi)$ is achieved by a prescription $\tilde\gamma: \mathcal{X} \times \mathcal{E} \mapsto [0,1]$ of the form:
\begin{align}
\tilde\gamma(x,e) = \left \{ \begin{array}{ll}
               1 & \mbox{if $|x-a| > n(e,\pi)$} \\
               0 & \mbox{ if $|x-a| < n(e,\pi)$} \\
               \alpha(e,\pi) & \mbox{if $x = a + n(e,\pi)$} \\
               \beta(e,\pi) & \mbox{if $x = a - n(e,\pi)$} 
                \end{array}
               \right. \label{eq:thresh}
          \end{align} 
         
%
where for each $e \in \mathcal{E}$, $\alpha(e,\pi),\beta(e,\pi) \in [0,1], \alpha(e,\pi) \leq \beta(e,\pi)$ and  $n(e,\pi)$ is a  non-negative integer.
\end{lemma}

\begin{proof} See Appendix \ref{sec:lemmaproof}.
  \end{proof}
Lemmas~\ref{lemma:est result} and  \ref{lemma:thresh1} can be used to establish a threshold structure for optimal prescriptions and a simple recursive optimal estimator for Problem \ref{prob:2}. At time $t=1$, by assumption A1, $\Pi_1$ is such that $\Pi_1(\cdot,e)$ is a.s.u. about $0$ for all $e \in \mathcal{E}$. Hence, by Lemma~\ref{lemma:thresh1}, an optimal prescription at time $t=1$ has the threshold structure of \eqref{eq:thresh}. If a transmission occurs at time $t=1$, then the resulting post-transmission belief $\Theta_1$ is a delta-function  and consequently $\Theta_1(\cdot,e), e \in \mathcal{E}$ are a.s.u. about the same point. If a transmission does not happen at time $t=1$, then, using Lemma \ref{lemma:update} and the threshold nature of  the prescription, it can be shown that the resulting post-transmission belief is such that $\Theta_1(\cdot,e), e \in \mathcal{E}$ are a.s.u. about $0$. Thus, it follows that $\Theta_1$ will always be such that all $\Theta_1(\cdot,e), e \in \mathcal{E}$ are a.s.u. about the same point and because of Lemma \ref{lemma:est result}, this point will be the optimal estimate. Using Lemma \ref{lemma:update} and the a.s.u. property of $\Theta_1(\cdot,e)$, it follows that the next pre-transmission belief $\Pi_2$ will always be such that $\Pi_2(\cdot,e), e \in \mathcal{E}$ are a.s.u. about the same point (by arguments similar to those in Lemma \ref{lemma:14} in Appendix \ref{sec:claimproof}). Hence, by Lemma~\ref{lemma:thresh1}, an optimal prescription at time $t=2$ has the threshold structure of \eqref{eq:thresh}. Proceeding sequentially as above establishes the following result.

\begin{theorem}\label{thm:2result}
In Problem \ref{prob:2}, under Assumptions A1 and A2, there is an optimal prescription and estimation strategy such that
\begin{enumerate}

\item The optimal estimate is given as:
\begin{align}
\hat X_t = \left \{ \begin{array}{ll}
               \hat X_{t-1} & \mbox{if $y_t =\epsilon$} \\
               x & \mbox{ if $y_t=(x,e)$ }
               \end{array},
               \right.
\end{align} 
where $\hat X_0 := 0$.
\item The pre-transmission belief at any time $t$, $\Pi_t(\cdot,e)$ is a.s.u. about $\hat X_{t-1}$, for all $e \in \mathcal{E}$.
\item The prescription at any time has the threshold structure of Lemma~\ref{lemma:thresh1}.
\end{enumerate}
\end{theorem}

As argued in Section \ref{sec:equiv prob}, Problem \ref{prob:2} and Problem \ref{prob:main} are equivalent. Hence, the result of Theorem~\ref{thm:2result} implies the following result for Problem \ref{prob:main}.

\begin{theorem}\label{thm:main}
 In Problem \ref{prob:main} under assumptions A1 and A2, there exist optimal decision strategies $\VEC f, \VEC g$  for the sensor and the estimator given as:
 
\begin{align}
g^*_t(y_{1:t}) = \left \{ \begin{array}{ll}
               a & \mbox{if $y_t =\epsilon$} \\
               x & \mbox{ if $y_t=(x,e)$ }
               \end{array}
               \right. \label{eq:main1}
\end{align} 
 
\begin{align}
f^*_t(x,e,y_{1:t-1}) = \left \{ \begin{array}{ll}
               1 & \mbox{if $|x-a| > n_t(e,\pi_t)$} \\
               0 & \mbox{ if $|x-a| < n_t(e,\pi_t)$} \\
               \alpha_t(e,\pi_t) & \mbox{if $x = a + n_t(e,\pi_t)$} \\
               \beta_t(e,\pi_t) & \mbox{if $x = a - n_t(e,\pi_t)$} 
                \end{array}
               \right. \label{eq:main2}
\end{align} 
where $a=0$ for $t=1$, $a = g^*_{t-1}(y_{1:t-1})$ for $t>1$, and  $\pi_t = \mathds{P}(X_t,E_t|y_{1:t-1})$.

\end{theorem}

Theorem \ref{thm:main} can be interpreted as follows: it says that the optimal estimate is  the most recently received value of the source (the optimal estimate is $0$ if no source value has been received). Further, there is a threshold rule at the sensor. The sensor transmits with probability 1 if the difference between the current source value and the most recently transmitted value exceeds a threshold that depends on sensor's current energy level and the estimator's pre-transmission belief; it does not transmit if the difference between the current source value and the most recently transmitted value is strictly below the  threshold.
\subsection{Optimal Thresholds}
 Theorem \ref{thm:main} gives a complete characterization of the optimal estimation strategy, but it only provides a \emph{structural form} of the optimal strategy for the sensor. Our goal  now is to find the exact characterization of the thresholds and the randomization probabilities in the structure of optimal strategy of the sensor. We  denote the optimal estimation strategy of Theorem \ref{thm:main} by $\mathbf{g}^*$ and the class of sensor strategies that satisfy the threshold structure of Theorem \ref{thm:main} as $\mathcal{F}$. We know that the global minimum expected cost is $J(\VEC f,\VEC g^*)$, for some $\VEC f \in \mathcal{F}$.
 Any sensor strategy $\VEC f'$ that achieves a cost $J(\VEC f', \VEC g^*) \leq J(\VEC f,\VEC g^*)$, for all $\VEC f \in \mathcal{F}$ must be a globally optimum sensor strategy.
 \par
 Given that the strategy for the estimator is fixed to $\VEC g^*$, we will address the question of finding the best sensor strategy among all possible strategies (including those not in $\mathcal{F}$). The answer to this question can be found by a standard dynamic program (see Lemma \ref{lemma:sdp} below). We denote  by $\VEC f^*$ the strategy specified by the dynamic program.  We have that $J(\VEC f,\VEC g^*) \geq J(\VEC f^*,\VEC g^*)$, for all $\VEC f$, (including those not in $\mathcal{F}$). Thus, $\VEC f^*$ is a globally optimal sensor strategy. Further, $\VEC f^*$ is in the set $\mathcal{F}$. Thus the dynamic program of Lemma \ref{lemma:sdp} provides a way of computing the optimal thresholds of Theorem \ref{thm:main}. 
 
\begin{lemma}\label{lemma:sdp}
Given that the strategy for the estimator is fixed to $\VEC g^*$,  the best sensor strategy (from the class of all possible strategies) is  of the form $ U_t = f^*_t(D_t,E_t), $
where $D_t := X_t-g^*_{t-1}(Y_{1:t-1})$. Further, this strategy is described by the following dynamic program:
\begin{equation*}
J_{T+1}(\cdot,\cdot) := 0
\end{equation*}
For positive energy levels $e > 0$,
\begin{align}\label{eq:sdpeq1}
&J_t(d,e) := \min\{c+\mathds{E}[J_{t+1}(Z_t,\min(e-1+N_t,B))], \notag \\ &\tilde{\rho}(d)+ \mathds{E}[J_{t+1}(d+Z_t,\min(e+N_t,B))] \},
\end{align}
where $\tilde{\rho}(d)$ is $\mathds{1}_{\{d\neq0\}}$ if the distortion metric is $\rho(x,a)=\mathds{1}_{\{x\neq a\}}$ and $\tilde{\rho}(d)$ is $|d|^k$ if the distortion metric is $\rho(x,a)=|x-a|^k$. 
For $e>0$, the optimal action for a realization $(d,e)$ of $(D_t,E_t)$ is $U_t=1$ iff $J_t(d,e)$ is equal to the first term in the right hand side of \eqref{eq:sdpeq1}.
If $e=0$, $J_t(\cdot,0) $ is the second term in the right hand side of \eqref{eq:sdpeq1} evaluated at $e=0$  and the optimal action is $U_t=0$.
\end{lemma} 
\begin{proof}
Once the estimator's strategy is fixed to $\VEC g^*$, the sensor's optimization problem is a standard Markov decision problem (MDP) with $D_t=X_t-g^*_{t-1}(Y_{1:t-1})$ and $E_t$ as the (two-dimensional) state. The result of the lemma is the  standard dynamic program for MDPs.
\end{proof}

Consider the definition of $J_t(d,e)$ in \eqref{eq:sdpeq1}. For a fixed $e>0$, the first term on right hand side of \eqref{eq:sdpeq1} does not depend on  $d$, while it can be easily shown that the second term is non-decreasing in $d$. These observations imply that for each $e>0$, there is a threshold value of $d$ below which $U_t=0$ and above which $U_t=1$ in the optimal strategy. Thus, the $\VEC f^*$ of Lemma \ref{lemma:sdp} satisfies the  threshold structure of Theorem \ref{thm:main}. Comparing the strategy $\VEC f^*$ specified by Lemma \ref{lemma:sdp} and the form of sensor strategies in Theorem \ref{thm:main}, we see that
\begin{enumerate}
\item  The  thresholds in $\VEC f^*$ depend \emph{only on the current energy level of the sensor} and not on the pre-transmission belief $\pi_t$ whereas the thresholds in Theorem \ref{thm:main} could depend on both energy level and $\pi_t$.
\item The strategy $\VEC f^*$ is purely deterministic whereas Theorem \ref{thm:main} allowed for possible randomizations at two points.
\end{enumerate}

\section{Multi-dimensional Gaussian Source} \label{sec:multi_d}

In this section, we consider a variant of Problem \ref{prob:main}, with a  multi-dimensional Gaussian source. The state of the source evolves according to the equation
\begin{equation}
\VEC X_{t+1} = \lambda\VEC A\VEC X_t + \VEC Z_t,
\end{equation}
where   $\VEC{X}_t =(X^1_t,X^2_t,\ldots,X^n_t)$, $\VEC{Z}_t =(Z^1_t,Z^2_t,\ldots,Z^n_t)$ are random vectors taking values in $\mathbb{R}^n$, $\lambda > 0$ is a real number and $\VEC A$ is an orthogonal matrix (that is, transpose of $\VEC A$ is the inverse of $\VEC A$ and, more importantly for our purpose, $\VEC A$ preserves norms). The initial state $\VEC X_1$ has a zero-mean Gaussian distribution with covariance matrix $s_1 \VEC I$, and $\VEC Z_1, \VEC Z_2,...,\VEC Z_{T-1}$ are i.i.d. random vectors with a zero-mean Gaussian distribution and covariance matrix $s_2 \VEC I$. The energy dynamics for the sensor are the same as in Problem \ref{prob:main}. 

 At the beginning of the time period $t$, the sensor makes a decision about whether to transmit its current observation vector and its current energy level to the estimator or not. 
The estimator receives a message $\VEC Y_t$ from the sensor where $\VEC Y_t = (\VEC{X}_t,E_t)$, if $U_t=1$ and $\VEC Y_t =\epsilon$ otherwise.
The estimator produces an estimate $\VEC{\hat{X}}_t = (\hat{X}^1_t,\ldots, \hat{X}^n_t)$ at time $t$ depending on the sequence of messages it received so far. The system operates for a finite time horizon $T$.

The sensor and estimator make their decisions according to deterministic strategies $\VEC f$ and $\VEC g$ of the form
$ U_t = f_t(\VEC X_t,E_t,\VEC Y_{1:t-1})$
and $\hat{\VEC{X}}_t = g_t(\VEC Y_{1:t})$. We assume that for any time and any realization of past messages, the set of source and energy states for which transmission happens is an open or a closed subset of $\mathbb{R}^n \times \mathcal{E}$. 
We have the following optimization problem.
\begin{problem}
  \label{prob:multi_d}
  For the model described above, given the statistics of the Markov source and the initial energy level $E_1$, the statistics of amounts of energy harvested at each time, the sensor's energy storage limit $B$ and the time horizon $T$,  find  decision strategies $\VEC f, \VEC g$ for the sensor and the estimator that minimize the following expected cost:
  \begin{equation}\label{eq:mcost}
  J(\VEC f, \VEC g) = \mathds{E} \{\sum_{t=1}^T c U_t + \norm{\VEC{X}_t-\VEC{\hat {X}}_t}^2 \},
  \end{equation}
 where $c \geq 0$ is a communication cost and $\norm{\cdot}$ is the Euclidean norm. 
\end{problem}
\begin{remark}
Note that we have assumed here that the sensor is using a deterministic strategy that employs only the current source and energy state and the past transmissions to make the decision at time $t$. Using arguments analogous to those used in proving Lemma \ref{lemma:str}, it can be shown that this restriction leads to no loss of optimality. While randomization was used in our proofs for the problem with discrete source (Problem \ref{prob:main}), it is not needed when the source state space is continuous.
\end{remark}



\begin{definition}
 A function $\nu: \mathbb{R}^n \mapsto \mathds{R}$  is said to be \emph{symmetric and unimodal} about a point $\VEC{a} \in \mathbb{R}^n$, if $\norm{\VEC{x-a}} \leq \norm{\VEC{y-a}}$ implies that $\nu(\VEC{x}) \geq \nu(\VEC{y})$. Further, we use the convention that a Dirac-delta function at $\VEC a$ is also symmetric unimodal about $\VEC a$.  
\end{definition}
For a Borel set $A$ in $\mathbb{R}^n$, we denote by $\mathcal{L}(A)$ the Lebesgue measure of $A$. 
\begin{definition}
For a Borel set $A$ in $\mathbb{R}^n$, we denote by $A^{\sigma}$ the symmetric rearrangement of $A$. That is, $A^{\sigma}$ is an open ball centered at $\VEC 0$ whose volume is $\mathcal{L}(A)$. Given an integrable, non-negative function $h: \mathbb{R}^n \mapsto \mathbb{R}$, we denote by $h^{\sigma}$ its symmetric non-decreasing rearrangement. That is, 
\[ h^{\sigma}(\VEC x) = \int_0^{\infty}\mathds{1}_{\{\VEC a \in \mathbb{R}^n|h(\VEC a) > t\}^{\sigma}}(\VEC x)dt \]
\end{definition}
\begin{definition}
 Given two integrable, non-negative functions $h_1$ and $h_2$ from $\mathbb{R}^n$ to $\mathbb{R}$, we say that $h_1$ majorizes $h_2$, denoted by, $h_2 \prec h_1$, if the following holds:
\begin{align}\label{eq:majdef}
\int_{\norm{\VEC x}\leq t}h_2^{\sigma}(\VEC x)d\VEC x \leq \int_{\norm{\VEC x}\leq t}h_1^{\sigma}(\VEC x)d\VEC x \mbox{~~~~$\forall t>0$}
\end{align} 
and
\begin{align*}
\int_{\mathbb{R}^n}h_2^{\sigma}(\VEC x)d\VEC x = \int_{\mathbb{R}^n}h_1^{\sigma}(\VEC x)d\VEC x 
\end{align*}
The condition in \eqref{eq:majdef} is equivalent to saying that for every Borel set $\mathbb{B} \subset \mathbb{R}^n$, there exists another Borel set $\mathbb{B}'\subset \mathbb{R}^n$ such that $\mathcal{L}(\mathbb{B}) = \mathcal{L}(\mathbb{B}')$ and $\int_{\mathbb{B}}h_2(\VEC x)d\VEC x \leq \int_{\mathbb{B}'}h_1(\VEC x)d\VEC x$.
\end{definition}

Following the arguments of Sections \ref{sec:prelim} and \ref{sec:results}, we can view the problem from the estimator's perspective who at each time $t$ selects a prescription for the sensor before the transmission and then an estimate on the source after the transmission. Because we have deterministic policies, the prescriptions are binary-valued functions. We can define at each time $t$,  the estimator's pre-transmission (post-transmission) beliefs as conditional probability densities on  $\mathbb{R}^n \times \mathcal{E}$ given the transmissions $\VEC Y_{1:t-1}$ ($\VEC Y_{1:t}$).
\begin{lemma}\label{lemma:Gupdate}
The estimator's beliefs evolve according to the following fixed transformations:
 \begin{enumerate}
\item $ \Pi_{t+1}(\VEC x,e) =\lambda^{-n}\int_{\VEC x' \in \mathbb{R}^n}\sum_{e' \in \mathcal{E}} [\mathds{P}(E_{t+1} =e|E'_t = e')\mu(\VEC x - \VEC x')  \Theta_t(\lambda^{-1}\VEC{A}^{-1}\VEC x',e')], $
where $\mu$ is the probability density function of $\VEC Z_t$. We denote this transformation by 
$\Pi_{t+1} = Q^1_{t+1}(\Theta_t)$.
\item 
\begin{align}
\Theta_t(\VEC x,e) =  \left \{ \begin{array}{ll}
               \delta_{\{\VEC x',e'-1\}}~~~ ~~~~~~~~~~~~ \mbox{if $\VEC Y_t =(\VEC x',e')$} \\
                \frac{(1-\Gamma_t(\VEC \VEC x,e))\Pi_t(\VEC x,e)}{\int_{\VEC x'}\sum_{e'}(1-\Gamma_t(\VEC x',e'))\Pi_t(\VEC x',e')}  \mbox{ if $\VEC Y_t=\epsilon$ }
               \end{array},
               \right.
\end{align} 
where $\delta_{\{\VEC x',e'-1\}}$ is a degenerate distribution  at $(\VEC x',e'-1)$.
 We denote this transformation by  $\Theta_t = Q^2_t(\Pi_t,\Gamma_t,\VEC Y_t)$. 
\end{enumerate}
\end{lemma}
Further, we can establish the following analogue of Theorem \ref{thm:val_functions} using dynamic programming arguments \cite{Shreve}.
\begin{theorem}\label{thm:mval_functions}
Let $\pi, \theta$ be any pre-transmission and post-transmission belief. Define recursively the following functions:
\begin{equation*}
W_{T+1}(\pi) := 0
\end{equation*}
\begin{align}\label{eq:mdpeq1}
&V_t(\theta) := \inf_{\VEC{a} \in \mathbb{R}^n} \mathds{E}[\norm{\VEC{X}_t - \VEC{a}}^2 + W_{t+1}(\Pi_{t+1})|\Theta_t=\theta] 
\end{align}
 where $\Pi_{t+1} = Q^1_{t+1}(\Theta_t)$ (see Lemma \ref{lemma:Gupdate}), and 
\begin{align} \label{eq:mdpeq2}
&W_{t}(\pi) := \inf_{\tilde\gamma \in \mathcal{G}} \mathds{E}[c\ind_{\{U_t=1\}}+V_{t}(\Theta_{t})|\Pi_t=\pi,\Gamma_t = \tilde\gamma] 
\end{align}
where $\mathcal{G}$ is the set of all functions $\gamma$ from $\mathbb{R}^n \times \mathcal{E}$ to $\{0,1\}$ such that $\gamma^{-1}(\{0\}) = \mathbb{R}^n \times \{0\} \cup (\cup_{e=1}^B \mathcal{I}_{e} \times \{e\})$, where $\mathcal{I}_e$ is an open or closed subset of $\mathbb{R}^n$.
\par
Then, $V_1(\pi_1)$, where $\pi_1$ is the density of $\VEC X_1$,  is a lower bound on the cost of any strategy; A strategy that at each time and for each realization of pre-transmission and post-transmission belief selects a prescription and an estimate that achieves the infima in \eqref{eq:mdpeq1} and \eqref{eq:mdpeq2} is optimal. Further, even if the infimum are not always achieved, it is possible to find a strategy with performance arbitrarily close to the lower bound. 
\end{theorem}

In order to completely characterize the solution of the dynamic program in Theorem \ref{thm:mval_functions}, we define the following relation on the possible realizations of estimator's beliefs.


\begin{definition} [Binary Relation $\RELATION^n$]
Let $\theta$ and $\tilde\theta$ be two post-transmission beliefs.
We say $\theta\RELATION\tilde\theta$ iff:
\begin{enumerate}
\item [(i)] For each $e \in \mathcal{E}$, 
$\theta(\cdot,e) \prec \tilde\theta(\cdot,e)$.
\item [(ii)] For all $e \in \mathcal{E}$, $\tilde\theta(\cdot,e)$ is symmetric and unimodal about the same point $ x \in \mathcal{X}$. 
\end{enumerate}
A similar relation holds for pre-transmission beliefs.
\end{definition}
\begin{definition}[Property $\RELATION^n$]
 Let $V$ be a function that maps probability measures on $\mathbb{R}^n\times\mathcal{E}$ to the set of real numbers $\mathbb{R}$. We say that $V$ satisfies Property $\RELATION^n$ iff for any two distributions $\theta$ and $\tilde\theta$,
 \[\theta \RELATION^n \tilde\theta \implies V(\theta) \geq V(\tilde\theta)\]
 
\end{definition}

We can now state the analogue of Claim \ref{claim:one}.

\begin{claim} \label{claim:two}
The value functions in Theorem \ref{thm:mval_functions}, $W_t$ $t=1,2,\ldots T+1$, and  $V_t, t=1,2,\ldots,T$, satisfy Property $\RELATION^n$.

\end{claim}
\begin{proof}
See Appendix \ref{sec:Gclaimproof}.
\end{proof}

Because of Claim 2, we can follow arguments similar to those in Section \ref{sec:results} to conclude the following: At time $t=1$, because $\pi_1(\cdot,e)$ is symmetric unimodal about $\VEC 0$ for all $e$, it is sufficient to consider symmetric threshold based prescriptions of the form 
\begin{align}
\gamma(\VEC x,e) = \left \{ \begin{array}{ll}
               1 & \mbox{if $\norm{\VEC x} \geq r_t(e,\pi_1)$} \\
               0 & \mbox{ if $\norm{\VEC x} < r_t(e,\pi_1)$}                
                \end{array}
               \right.
\end{align} 
in right hand side of equation \eqref{eq:mdpeq1} for time $t=1$. Using such prescriptions implies that $\theta_1(\cdot,e)$ is always symmetric unimodal about some point $\VEC a$ which is the optimal estimate in \eqref{eq:mdpeq2} at time $t=1$. Further, $\pi_2(\cdot,e)$ will also be symmetric unimodal about $\lambda \VEC A \VEC a$ and therefore it is sufficient to restrict to symmetric threshold based prescriptions in \eqref{eq:mdpeq2} at time $t=2$. Proceeding sequentially till time $T$ allows us to conclude that at each time, we only need to consider pre and post transmission beliefs that are symmetric unimodal, prescriptions that are symmetric threshold based and estimates that are equal to the point about which the belief is symmetric. Then, we can conclude the following result. 
\begin{theorem}\label{thm:Gmain}
 In Problem \ref{prob:multi_d}, it is without loss of optimality\footnote{That is, there is a strategy of the form in the theorem whose performance is arbitrarily close to the lower bound $V_1(\pi_1)$} to restrict to  strategies  $\VEC f^*, \VEC g^*$  that are given as:
\begin{align}
g^*_t(\VEC y_{1:t}) = \left \{ \begin{array}{ll}
                \lambda \VEC{A}\VEC{a} & \mbox{if $\VEC y_t =\epsilon$} \\
               \VEC x & \mbox{ if $\VEC y_t=(\VEC x,e)$ }
               \end{array}
               \right.
\end{align} 
 
\begin{align}
f^*_t(\VEC x,e,\VEC y_{1:t-1}) = \left \{ \begin{array}{ll}
               1 & \mbox{if $\norm{\VEC x- \lambda \VEC{A}\VEC a} \geq r_t(e,\pi_t)$} \\
               0 & \mbox{ if $\norm{\VEC x- \lambda \VEC{A}\VEC a} < r_t(e,\pi_t)$}                
                \end{array}
               \right.
\end{align} 
where  $\VEC a=\VEC 0$ for $t=1$, $\VEC a = g^*_{t-1}(\VEC y_{1:t-1})$ for $t>1$, $\pi_t = \mathds{P}(X_t,E_t|\VEC y_{1:t-1})$, and $r_t(e,\pi_t) \geq 0$.
\end{theorem}
Further, the optimal values of thresholds can be obtained by the following dynamic program which is similar to the dynamic program in  Lemma \ref{lemma:sdp}.
\begin{lemma}\label{lemma:Gsdp}
Given that the strategy for the estimator is fixed to $\VEC g^*$,  the best sensor strategy (from the class of all possible strategies) is  of the form $ U_t = f^*_t(\VEC D_t,E_t), $
where $\VEC D_t := \VEC X_t-g^*_{t-1}(Y_{1:t-1})$. Further, this strategy is described by the following dynamic program:
\begin{equation*}
J_{T+1}(\cdot,\cdot) := 0
\end{equation*}
For positive energy levels $e > 0$,
\begin{align}\label{eq:Gsdpeq1}
&J_t(\VEC d,e) := \min\{c+\mathds{E}[J_{t+1}(\VEC Z_t,\min(e-1+N_t,B))], \notag \\ &\norm{\VEC d}^2+ \mathds{E}[J_{t+1}(\VEC d+\VEC Z_t,\min(e+N_t,B))] \},
\end{align}
For $e>0$, the optimal action for a realization $(\VEC d,e)$ of $(\VEC D_t,E_t)$ is $U_t=1$ iff $J_t(\VEC d,e)$ is equal to the first term in the right hand side of \eqref{eq:Gsdpeq1}.
If $e=0$, $J_t(\cdot,0) $ is the second term in the right hand side of \eqref{eq:Gsdpeq1} evaluated at $e=0$  and the optimal action is $U_t=0$.
\end{lemma}

\section{Special Cases}\label{sec:special}
By making suitable assumptions on the source, the energy storage limit $B$ of the sensor and statistics of initial energy level and the energy harvested at each time, we can derive the following special cases of Problem \ref{prob:main} in Section \ref{sec:PF} and Problem \ref{prob:multi_d} in Section \ref{sec:multi_d}.
\subsubsection{Fixed number of Transmissions}  Assume that the initial energy level $E_1 = K$ ($K \leq B$) with probability $1$ and that the energy harvested at any time is $N_t =0$ with probability $1$. Under these assumptions, Problem \ref{prob:main} can be interpreted as capturing the scenario when the sensor can afford at most $K$ transmissions during the time-horizon with no possibility of energy harvesting. This is similar to the model in \cite{ImerBasar1}.

\subsubsection{No Energy Constraint} Assume that the storage limit $B=1$ and that initial energy level and the energy harvested at each time is $1$ with probability $1$. Then, it follows that at any time $t$, $E_t=1$ with probability $1$. Thus, the sensor is \emph{always guaranteed to have energy to communicate}. Under these assumptions, Problem \ref{prob:main} can be interpreted as capturing he scenario when the sensor has no energy constraints (it still has energy costs because of the term $cU_t$ in the objective). This is similar to the model in \cite{Martins:2011}.

\subsubsection{I.I.D. Source} The analysis of Sections \ref{sec:results} and \ref{sec:multi_d} can be repeated if the source evolution is assumed to be $X_{t+1} = Z_t$, where $Z_t$ are the i.i.d. noise variables. For i.i.d. sources, the optimal estimate is the mean value of the source in case of no transmission.  Also, the dynamic program of Lemma \ref{lemma:sdp} can be used for finite valued i.i.d. sources by replacing $D_t$ with $X_t$ and changing \eqref{eq:sdpeq1} to
$J_t(d,e) := \min\{c+\mathds{E}[J_{t+1}(X_{t+1},\min(e-1+N_t,B))], \tilde{\rho}(d)+ \mathds{E}[J_{t+1}({X}_{t+1},\min(e+N_t,B))] \}$.
A similar dynamic program can be written for the Gaussian source.

\section{Conclusion}\label{sec:conc}
We considered the problem of finding globally optimal communication scheduling and estimation strategies in a remote estimation problem with an energy harvesting sensor and a finite-valued or a multi-dimensional Gaussian source. We established the global optimality of a simple energy-dependent threshold-based  communication strategy and a simple estimation strategy. Our results considerably simplify the off-line computation of optimal strategies as well as their on-line implementation.
\par
Our approach started with providing a POMDP based dynamic program for the decentralized decision making problem. Dynamic programming solutions often rely on finding a key property of value functions (such as concavity or quadratic-ness) and exploiting this property to characterize the solution. In dynamic programs that arise from decentralized problems, however, value functions involve minimization over functions \cite{mythesis} and hence the usual properties of value functions are either not applicable or not useful.  In such problems, there is a need to find the right property of value functions that can be used to characterize optimal solutions. We believe that this work demonstrates that, in some problems, majorization based properties related to Schur concavity may be the right value function property to exploit. 
\section{Acknowledgments}
This work was supported in part by NSF under grant numbers CCF 11-11342 and CCF 11-11061 and by NASA Grant NNX06AD47G.

\appendices

\section{Lemmas from \cite{Hajek:2008}, Section VI}
\subsection{For the discrete source}
\begin{lemma}\label{lemma:6.2}
If $\mu$ is a.s.u. and even  and $\xi$ is a.s.u. about $a$, then the convolution $\xi \conv \mu$ is a.s.u about $a$.
\end{lemma}
\begin{lemma}\label{lemma:6.3}
If $\mu$ is a.s.u. and even, $\tilde{\xi}$ is a.s.u. and $\xi \prec \tilde\xi$, then $\xi\conv\mu \prec \tilde{\xi}\conv\mu$.
\end{lemma}
\subsection{For the multi-dimensional Gaussian source}
\begin{lemma}\label{lemma:10}
If $\mu$ and $\nu$ are two non-negative integrable functions on $\mathbb{R}^n$ and $\mu \prec \nu$, then $\int_{\mathbb{R}^n} \mu^{\sigma}(\VEC x)h(\VEC x) \leq \int_{\mathbb{R}^n} \nu^{\sigma}(\VEC x)h(\VEC x)$ for any symmetric unimodal function $h$.
\end{lemma}
\begin{lemma}\label{lemma:11}
If $\mu$ and $\nu$ are two non-negative integrable functions on $\mathbb{R}^n$, then $\int_{\mathbb{R}^n} \mu(\VEC x)\nu(\VEC x) \leq \int_{\mathbb{R}^n} \mu^{\sigma}(\VEC x)\nu^{\sigma}(\VEC x)$ (This lemma is known as the Hardy Littlewood Inequality \cite{hardy}).
\end{lemma}

\begin{lemma}\label{lemma:6.7}
If $\mu$ is symmetric unimodal about $\VEC 0$, $\tilde{\xi}$ is symmetric unimodal and $\xi \prec \tilde\xi$, then $\xi\conv\mu \prec \tilde{\xi}\conv\mu$.
\end{lemma}
\section{Other Preliminary Lemmas} \label{sec:prelemmas}

\begin{lemma}\label{lemma:new}
Let $h_1$ be a non-negative, integrable functions from $\mathbb{R}^n$ to $\mathbb{R}$ such that $h_1$ is symmetric unimodal about a point $\VEC a$. Let $h_2$ be a pdf on $\mathbb{R}^n$ that is symmetric unimodal  about $\VEC 0$. Then, $h_1 \conv h_2$ is symmetric unimodal about $\VEC a$.
\end{lemma}

\begin{proof}
For ease of exposition, we will assume that both $h_1$ and $h_2$ are symmetric unimodal about $\VEC 0$. If $h_1$ is symmetric unimodal about a non-zero point, then to obtain $h_1 \conv h_2$ we can first do a translation of $h_1$ so that it is symmetric unimodal about $\VEC 0$, carry out the convolution and translate the result back.
\par
Consider two points $\VEC x \neq \VEC y$ such that $\norm{\VEC x}=\norm{\VEC y}$. Then, we can always find an orthogonal matrix such that $\VEC y= Q \VEC x$. Then,
\begin{align}
&(h_1\conv h_2)(\VEC y) = (h_1\conv h_2)(Q\VEC x) = \int_{\VEC z}h_1(\VEC z)h_2(Q \VEC x - \VEC z)d\VEC z 
\end{align}
Carrying out a change of variables so that $\VEC z = Q\VEC z'$, the above integral becomes
\begin{align}
&\int_{\VEC z'}h_1(Q\VEC z')h_2(Q \VEC x - Q\VEC z')d \VEC z' = \int_{\VEC z'}h_1(Q\VEC z')h_2(Q (\VEC x - \VEC z'))d \VEC z' \notag \\
&=\int_{\VEC z'}h_1(\VEC z')h_2(\VEC x - \VEC z')d \VEC z' = (h_1\conv h_2)(\VEC x)
\end{align}
where we used the symmetric nature of $h_1$ and $h_2$ and the fact that the orthogonal matrix preserves norm. Thus, any two points with the same norm have the same value of $h_1 \conv h_2$. This establishes the symmetry of $h_1 \conv h_2$. Next, we look at unimodality.
We follow an argument similar to the one used in \cite{Purkayastha}.
Because of symmetry, it suffices to show that $(h_1 \conv h_2)(x_1,0,0...,0)$ is non-increasing for $x_1 \in [0,\infty)$. (Here, $(x_1,0,...,0)$ is the $n$ dimensional vector with all but the first coordinates as $0$.)
\begin{align}
(h_1 \conv h_2)((x_1,0,...,0)) = \int_{\VEC z}h_2(\VEC z)h_1((x_1,0..,0) - \VEC z)d\VEC z = \mathds{E}[h_1((x_1,0..,0) - \VEC Z)],
\end{align}
where $\VEC Z$ is a random vector with pdf $h_2$. Define a new random variable $Y_{x_1} := h_1((x_1,0..,0) - \VEC Z)$. Then,
\begin{align}
\mathds{E}[h_1((x_1,0..,0) - \VEC Z)] = \mathds{E}[Y_{x_1}] = \int_{0}^{\infty}\mathds{P}(Y_{x_1} > t)dt \label{eq:nconv}
\end{align}
We now prove that for any given $t \geq 0$, $\mathds{P}(Y_{x_1} > t)$ is non-increasing in $x_1$. This would imply that the integral in \eqref{eq:nconv}  and hence $(h_1 \conv h_2)((x_1,0...0))$ is non-increasing in $x_1$.
\par
 The symmetric unimodal nature of $h_1$ implies that $Y_{x_1} > t$ if and only if $\norm{(x_1,0..,0) - \VEC Z} < r$ (or $\norm{(x_1,0..,0) - \VEC Z} \leq r$) for some constant $r$ whose value varies with $t$.
Thus,
\begin{align}
&\mathds{P}(Y_{x_1} > t) = \mathds{P}(\norm{(x_1,0..,0) - \VEC Z} < r)  = \int_{\mathbb{S}(x_1,r)}h_2(\VEC z) d \VEC z, \label{eq:nconv2}
\end{align}
where $\mathbb{S}(x_1,r)$ is the $n$-dimensional (open) sphere centered at $(x_1,0,..0)$ with radius $r$. It can be easily verified that the symmetric unimodal nature of $h_2$ implies that as the center of the sphere $\mathbb{S}(x_1,r)$ is shifted away from the origin (keeping the radius fixed), the integral in \eqref{eq:nconv2} cannot increase. This concludes the proof.
\end{proof}

\section{Proof of Claim 1}\label{sec:claimproof}
Since $W_{T+1}(\pi):=0$ for any choice of $\pi$, it trivially satisfies Property $\RELATION$. We will now proceed in a backward inductive manner.

\textbf{Step 1:} If $W_{t+1}$ satisfies Property $\RELATION$, we will show that $V_t$ satisfies Property $\RELATION$ too. \\
Using Lemma \ref{lemma:update}, the expression in \eqref{eq:dpeq1} can be written as
\begin{align}
 V_t(\theta) := W_{t+1}(Q^1_{t+1}(\theta)) + \min_{a \in \mathcal{X}} \mathds{E}[\rho(X_t, a) |\Theta_t=\theta] \label{eq:claimeq1}
 \end{align}
We will look at the two terms in the above expression separately and show that each term satisfies Property $\RELATION$. To do so, we will use the following lemmas.
\begin{lemma} \label{lemma:14}
$\theta \RELATION \tilde\theta \implies Q^1_{t+1}(\theta) \RELATION Q^1_{t+1}(\tilde\theta). $
\end{lemma}
\begin{proof}
Let $\pi = Q^1_{t+1}(\theta)$ and $\tilde\pi = Q^1_{t+1}(\tilde\theta)$.
Then, from Lemma \ref{lemma:update}, 
\begin{align}
 &\pi(x,e) = \sum_{\substack{x' \in \mathbb{Z}, \\e' \in \mathcal{E}}} [\mathds{P}(X_{t+1} =x|X_t = x') \mathds{P}(E_{t+1} =e|E'_t = e') \theta_t(x',e')] \notag \\
 &=\sum_{e' \in \mathcal{E}}\Big[\mathds{P}(E_{t+1} =e|E'_t = e') \sum_{x' \in \mathbb{Z}}[\mathds{P}(X_{t+1} =x|X_t = x')\theta(x',e')] \Big] \notag 
 \end{align}
 \begin{align}
 &=\sum_{e' \in \mathcal{E}}\Big[\mathds{P}(E_{t+1} =e|E'_t = e') \sum_{x' \in \mathbb{Z}}[\mathds{P}(Z_t =x-x')\theta(x',e')] \Big] \notag \\
 &=\sum_{e' \in \mathcal{E}}\mathds{P}(E_{t+1} =e|E'_t = e')\zeta(x,e'), 
 \end{align}
 where $\zeta(x,e') = \sum_{x' \in \mathbb{Z}}[\mathds{P}(Z_t =x-x')\theta(x',e')]$.
 Similarly, 
 \begin{align}
 \tilde\pi(x,e) = \sum_{e' \in \mathcal{E}}\mathds{P}(E_{t+1} =e|E'_t = e')\tilde\zeta(x,e')
 \end{align}
 where $\tilde\zeta(x,e') = \sum_{x' \in \mathbb{Z}}[\mathds{P}(Z_t =x-x')\tilde\theta(x',e')]$. 
 In order to show that $\pi(\cdot,e) \prec \tilde\pi(\cdot,e)$, it suffices to show that $\zeta(\cdot,e') \prec \tilde{\zeta}(\cdot,e')$ and that $\tilde{\zeta}(\cdot,e')$ are a.s.u about the same point for all $e' \in \mathcal{E}$. It is clear that 
 \[ \zeta(\cdot,e') = \mu \conv \theta(\cdot,e'), ~~~~ \tilde{\zeta}(\cdot,e') = \mu \conv \tilde{\theta}(\cdot,e') \]
 where $\mu$ is the distribution of $Z_t$ and $\conv$ denotes convolution. We now use the result in Lemmas \ref{lemma:6.2} and \ref{lemma:6.3} from Appendix A to conclude that  $\mu \conv \theta(\cdot,e') \prec  \mu \conv \tilde{\theta}(\cdot,e')$ and that $\mu \conv \tilde{\theta}(\cdot,e')$ is a.s.u. about the same point as $\tilde{\theta}(\cdot,e')$. Thus, we have established that for all $e \in \mathcal{E}$, $\pi(\cdot, e) \prec \tilde{\pi}(\cdot,e)$. 
  Similarly, we can argue that  $\tilde{\pi}(\cdot,e)$ are a.s.u. about the same point since $\tilde{\zeta}(\cdot,e')$ are a.s.u about the same point. 
 Thus,  \[\theta \RELATION \tilde\theta \implies Q^1_{t+1}(\theta) \RELATION Q^1_{t+1}(\tilde\theta). \]
 \end{proof}
The above relation combined with the assumption that $W_{t+1}$ satisfies Property $\RELATION$ implies that the first term in \eqref{eq:claimeq1} satisfies Property $\RELATION$.  The following lemma addresses the second term in \eqref{eq:claimeq1}. 
\begin{lemma}
Define $L(\theta) :=  \min_{a \in \mathcal{X}} \mathds{E}[\rho(X_t, a) |\Theta_t=\theta]$. $L(\cdot)$ satisfies Property $\RELATION$.
\end{lemma}
\begin{proof}
For any $a \in \mathcal{X}$, the conditional expectation in the definition of $L(\theta)$ can be written as
\begin{align}
&\sum_{x \in \mathbb{Z}}\rho(x,a)\Big\{\sum_{e \in \mathcal{E}}\theta(x,e)\Big\} \notag \\
&=\sum_{x \in \mathbb{Z}}\rho(x,a)m_{X}\theta(x) \label{eq:distort1}
\end{align} 
where $m_{X}\theta(x) = \sum_{e \in \mathcal{E}}\theta(x,e)$ is the marginal distribution of $\theta$. 
Recall that the distortion function $\rho(x,a)$ is a non-decreasing function of $|x-a|$. Let $d_i$ be the value of the distortion when $|x-a|=i$ Let $\mathcal{D} : = \{0,d_1,d_1,d_2,d_2,d_3,d_3,\ldots,d_M,d_M\}$, where $M$ is the cardinality of $\mathcal{X}$. It is clear that the expression in \eqref{eq:distort1} is an inner product of some permutation of $\mathcal{D}$ with $m_X\theta$. For any choice of $a$, such an inner product is lower bounded as
\begin{equation}
\sum_{x \in \mathcal{X}}\rho(x,a)m_{X}\theta(x) \geq \langle\mathcal{D}_{\uparrow}, m_{X}\theta_{\downarrow}\rangle,
\end{equation}
which implies that 
\begin{equation}
L(\theta) \geq \langle\mathcal{D}_{\uparrow}, m_{X}\theta_{\downarrow}\rangle, \label{eq:distort2}
\end{equation}
where $\langle\cdot,\cdot\rangle$ represents inner product, $\mathcal{D}_{\uparrow}$ is the non-decreasing rearrangement of $\mathcal{D}$ and $m_{X}\theta_{\downarrow}$ is the non-increasing rearrangement of $m_{X}\theta$. If $\theta \RELATION \tilde\theta$, then it follows that $m_X\theta \prec m_X\tilde\theta$ and $m_X\tilde\theta$ is a.s.u. about some point $b \in \mathcal{X}$. It can be easily established that $m_X\theta \prec m_X\tilde\theta$ implies that 
\begin{equation}\label{eq:distort2.1}
 \langle\mathcal{D}_{\uparrow}, m_{X}\theta_{\downarrow}\rangle \geq \langle\mathcal{D}_{\uparrow}, m_{X}\tilde\theta_{\downarrow}\rangle
 \end{equation}
 Further, since $m_X\tilde\theta$ is a.s.u. about $b$, $\sum_{x \in \mathcal{X}}\rho(x,b)m_{X}\tilde\theta(x) = \langle\mathcal{D}_{\uparrow}, m_{X}\tilde\theta_{\downarrow}\rangle$. Thus, 
 \begin{align}
 L(\tilde\theta) = \langle\mathcal{D}_{\uparrow}, m_{X}\tilde\theta_{\downarrow}\rangle \label{eq:distort3}
 \end{align}
 Combining \eqref{eq:distort2}, \eqref{eq:distort2.1} and \eqref{eq:distort3} proves the lemma.
  \end{proof}
Thus, both terms in \eqref{eq:claimeq1} satisfy Property $\RELATION$ and hence $V_t$ satisfies Property $\RELATION$.

\textbf{Step 2:} If $V_{t}$ satisfies Property $\RELATION$, we will show that $W_t$ satisfies Property $\RELATION$ too. \\
Consider two distributions $\pi$ and $\tilde\pi$ such that $\pi \RELATION \tilde \pi$. Recall that \eqref{eq:dpeq2} defined $W_t(\pi)$ as
\begin{align}
&W_{t}(\pi) =  \min_{\hat\gamma}\mathds{E}[c\ind_{\{U_t=1\}}+V_{t}(\Theta_{t})|\Pi_t=\pi,\gamma_t = \hat\gamma]  =: \min_{\hat\gamma} \mathds{W}(\pi,\hat\gamma) \label{eq:claimeq2}
\end{align}
where $\mathds{W}(\pi,\hat\gamma)$ denotes the conditional expectation in \eqref{eq:claimeq2}. 
Suppose that the  minimum in the definition of $W_t(\pi)$ is achieved by some prescription $\gamma$, that is, $W_t(\pi) = \mathds{W}(\pi,\gamma)$. Using $\gamma$,  we will construct another prescription $\tilde\gamma$ such that $\mathds{W}(\tilde\pi,\tilde\gamma) \leq \mathds{W}(\pi,\gamma)$. This will imply that $W_t(\tilde\pi) \leq W_t(\pi)$, thus establishing the statement of step 2. We start with 
   \begin{align}
&\mathds{W}(\pi,\gamma) =  \mathds{E}[c\ind_{\{U_t=1\}}+V_{t}(\Theta_{t})|\Pi_t=\pi,\gamma_t = \gamma] \notag\\
&= c\mathds{P}(U_t=1|\Pi_t=\pi,\gamma_t = \gamma) + \mathds{E}[V_{t}(\Theta_{t})|\Pi_t=\pi,\gamma_t = \gamma] \notag\\
&= c\sum_{x,e}\pi(x,e)\gamma(x,e) + \mathds{E}[V_{t}(Q^2_t(\pi,Y_t,\gamma))|\Pi_t=\pi,\gamma_t = \gamma]  \label{eq:dpeq2aux1}
\end{align}
The second term in \eqref{eq:dpeq2aux1} can be further written as
\begin{align} 
&\mathds{P}(Y_t=\epsilon|\Pi_t=\pi,\gamma_t = \gamma) \times 
[V_{t}(Q^2_t(\pi,Y_t=\epsilon,\gamma))|\Pi_t=\pi,\gamma_t = \gamma] + \notag\\
&\sum_{x,e}[\mathds{P}(Y_t=(x,e)|\Pi_t=\pi,\gamma_t = \gamma) \times [V_{t}(Q^2_t(\pi,Y_t=(x,e),\gamma))|\Pi_t=\pi,\gamma_t = \gamma]] \notag \\
&=\sum_{x',e'}\pi(x',e')(1-\gamma(x',e')) \times V_t(\theta^{\gamma}) 
+\sum_{x,e}\pi(x,e)\gamma(x,e)V_t(\delta(x,e-1)) \label{eq:depeq2aux2}
\end{align}
where $\theta^{\gamma}$ is the distribution resulting from $\pi$ and $\gamma$ when $Y_t =\epsilon$ (see Lemma \ref{lemma:update}). Substituting \eqref{eq:depeq2aux2} in \eqref{eq:dpeq2aux1} gives the minimum value to be
\begin{align}
&c\sum_{x,e}\pi(x,e)\gamma(x,e) + \sum_{x,e}\pi(x,e)\gamma(x,e)V_t(\delta_{(x,e-1)}) 
+ \sum_{x',e'}\pi(x',e')(1-\gamma(x',e')) \times V_t(\theta^{\gamma}) \label{eq:dpeq2aux3}
\end{align}

We will now use the fact that $V_t$ satisfies Property $\RELATION$ to conclude that $V_t(\delta_{(x,e-1)})$ does not depend on $x$. That is, $V_t(\delta_{(x,e-1)}) = K(e-1), \forall x \in \mathcal{X},$
where $K(e-1)$ is a number that only depends on $e-1$.
Consider $\delta_{(x,e-1)}$ and $\delta_{(x',e-1)}$. It is easy to see that
$\delta_{(x,e-1)}\RELATION\delta_{(x',e-1)}$ and  $\delta_{(x',e-1)}\RELATION\delta_{(x,e-1)}$.
Since $V_t$ satisfies Property $\RELATION$, it implies that $V_t(\delta_{(x',e-1)}) \leq V_t(\delta_{(x,e-1)})$ and $V_t(\delta_{(x,e-1)})\leq V_t(\delta_{(x',e-1)})$. Thus, $V_t(\delta_{(x,e-1)}) = V_t(\delta_{(x',e-1)}) = K(e-1)$. Equation~\eqref{eq:dpeq2aux3} now becomes
\begin{align}
&c\sum_{x,e}\pi(x,e)\gamma(x,e) + \sum_{x,e}\pi(x,e)\gamma(x,e)K(e-1) \notag \\
&+ \sum_{x',e'}\pi(x',e')(1-\gamma(x',e')) \times V_t(\theta^{\gamma}) \label{eq:dpeq2aux4}
\end{align}
We define $\lambda(e):= \sum_{x \in \mathcal{X}}\pi(x,e)(1-\gamma(x,e))$.

%
%
We will now construct another prescription $\tilde\gamma$. For that matter, we first define the sequence $\mathcal{S}=\{0,1,-1,2,-2,3,-3,\ldots...\}$ and let $s(n)$ denote the $n^{th}$ element of this sequence. Recall that $\tilde\pi(\cdot,e)$ is a.s.u. about the same point $a \in \mathcal{X}$ for all $e \in \mathcal{E}$.
For each $e \in \mathcal{E}$, define
\[n^*(e) := \min\{n: \sum_{k=1}^n \tilde\pi(a+s(k),e) \geq \lambda(e) \} \]
and
\[ \alpha(e) := \frac{\lambda(e) - \sum_{k=1}^{n^*(e)-1} \tilde\pi(a+s(k),e)}{\tilde\pi(a+s(n^*(e)),e)}.\]
Define $\tilde\gamma(\cdot,\cdot)$ as
\begin{align}
\tilde\gamma(a+s(k),e) = \left \{ \begin{array}{ll}
               0 & \mbox{if $k < n^*(e)$} \\
               (1-\alpha(e)) & \mbox{ if $k = n^*(e)$} \\
              1 & \mbox{if $k > n^*(e)$} 
                \end{array}
               \right. \label{eq:threshextra}
          \end{align}

          We can show that with the above choice of $\tilde\gamma$, 
          \begin{equation} \sum_{x}\pi(x,e)(1-\gamma(x,e)) = \sum_{x}\tilde\pi(x,e)(1-\tilde\gamma(x,e)).\label{eq:equality2}
          \end{equation}
          and
          \begin{equation} \sum_{x}\pi(x,e)\gamma(x,e) = \sum_{x}\tilde\pi(x,e)\tilde\gamma(x,e).\label{eq:equality}
          \end{equation}
        Using the same analysis used to obtain \eqref{eq:dpeq2aux4},  we can now evaluate the expression\[ \mathds{E}[c\ind_{\{U_t=1\}}+V_{t}(\Theta_{t})|\Pi_t=\tilde\pi,\gamma_t = \tilde\gamma] \] to be
        \begin{align}
        &c\sum_{x,e}\tilde\pi(x,e)\tilde\gamma(x,e) + \sum_{x,e}\tilde\pi(x,e)\tilde\gamma(x,e)K(e-1) + \sum_{x',e'}\tilde\pi(x',e')(1-\tilde\gamma(x',e')) \times V_t(\tilde\theta^{\tilde\gamma}), \label{eq:simplify1}
        \end{align}
where $\tilde\theta^{\gamma}$ is the distribution resulting from $\tilde\pi$ and $\tilde\gamma$ when $Y_t =\epsilon$ (see Lemma \ref{lemma:update}). Using \eqref{eq:equality} in \eqref{eq:simplify1}, we obtain the expression
      \begin{align}
        &c\sum_{x,e}\pi(x,e)\gamma(x,e) + \sum_{x,e}\pi(x,e)\gamma(x,e)K(e-1) \notag \\
&+ \sum_{x',e'}\pi(x',e')(1-\gamma(x',e')) \times V_t(\tilde\theta^{\tilde\gamma}), \label{eq:simplify2}
        \end{align}
        Comparing \eqref{eq:dpeq2aux4} and \eqref{eq:simplify2}, we observe that all terms in the two expressions are identical except for the last term $V_t(\cdot)$. Using the expressions for $\theta^{\gamma}$ and $\tilde\theta^{\tilde\gamma}$ from Lemma \ref{lemma:update} and the fact that $\pi \RELATION \tilde\pi$, it can be shown that $\theta^{\gamma} \RELATION \tilde\theta^{\tilde\gamma}$. Thus,  $V_t(\tilde\theta^{\tilde\gamma}) \leq V_t(\theta^{\gamma})$. This implies that the expression in \eqref{eq:simplify2} is no more than the expression in \eqref{eq:dpeq2aux4}. This establishes the statement of Step 2.

\section{Proof of Lemma \ref{lemma:thresh1}} \label{sec:lemmaproof}
Suppose that the  minimum in the definition of $W_t(\pi)$ is achieved by some prescription $\gamma$. Using $\gamma$,  we will construct another prescription $\tilde\gamma$ of the form in \eqref{eq:thresh} which also achieves the minimum. The construction of $\tilde \gamma$ is identical to the construction of $\tilde \gamma$ in Step 2 of the proof of Claim 1 (using $\pi$ instead of $\tilde \pi$ to define $n^*(e),\alpha(e)$). The a.s.u. assumption of $\pi$ and the nature of constructed $\tilde\gamma$ imply that $\tilde\gamma$ is of the form required in the Lemma. 

\section{Proof of Claim 2}\label{sec:Gclaimproof}
The proof follows a backward inductive argument similar to the proof of Claim 1.
\medskip\\
\textbf{Step 1:} If $W_{t+1}$ satisfies Property $\RELATION^n$, we will show that $V_t$ satisfies Property $\RELATION^n$ too. \\
Using Lemma \ref{lemma:Gupdate}, the expression in \eqref{eq:mdpeq1} can be written as
\begin{align}
 V_t(\theta) := W_{t+1}(Q^1_{t+1}(\theta)) + \inf_{\VEC a \in \mathbb{R}^n} \mathds{E}[\rho(\VEC X_t, \VEC a) |\Theta_t=\theta] \label{eq:mclaimeq1}
 \end{align}
 We will look at the two terms in the above expression separately and show that each term satisfies Property $\RELATION^n$.
 \begin{lemma}
$\theta \RELATION^n \tilde\theta \implies Q^1_{t+1}(\theta) \RELATION^n Q^1_{t+1}(\tilde\theta). $
\end{lemma}
\begin{proof}
Let $\pi = Q^1_{t+1}(\theta)$ and $\tilde\pi = Q^1_{t+1}(\tilde\theta)$.
Then, following steps similar to those in proof of Claim 1, 
\begin{align}
 \pi(\VEC{x},e) = \sum_{e' \in \mathcal{E}}\mathds{P}(E_{t+1} =e|E'_t = e')\zeta(\VEC{x},e'), 
 \end{align}
 where $\zeta(\VEC{x},e') = \lambda^{-n}\int_{\VEC{x}' \in \mathbb{R}^n}[\mu(\VEC{x-x'})\theta(\lambda^{-1}\VEC{A}^{-1}\VEC{x'},e')]$.
 Similarly, 
 \begin{align}
 \tilde\pi(\VEC x,e) = \sum_{e' \in \mathcal{E}}\mathds{P}(E_{t+1} =e|E'_t = e')\tilde\zeta(\VEC x,e')
 \end{align}
 where $\tilde\zeta(\VEC x,e') = \lambda^{-n}\int_{\VEC x' \in \mathbb{R}^n}[\mu(\VEC{x-x'})\tilde\theta(\lambda^{-1}\VEC{A}^{-1}\VEC{x'},e')]$.
 In order to show that $\pi(\cdot,e) \prec \tilde\pi(\cdot,e)$, it suffices to show that $\lambda^{n}\zeta(\cdot,e') \prec \lambda^{n}\tilde{\zeta}(\cdot,e')$ and that $\tilde{\zeta}(\cdot,e')$ are symmetric unimodal about the same point for all $e' \in \mathcal{E}$. It is clear that 
 \[ \lambda^{n}\zeta(\cdot,e') = \mu \conv \eta(\cdot,e'), ~~~~ \lambda^{n}\tilde{\zeta}(\cdot,e') = \mu \conv \tilde{\eta}(\cdot,e') \]
 where  $\eta(\VEC x,e') = \theta(\lambda^{-1}\VEC{A}^{-1}\VEC{x},e')$ and $\tilde\eta(\VEC x,e') = \tilde\theta(\lambda^{-1}\VEC{A}^{-1}\VEC{x},e')$. 
 Recall that $\theta(\cdot,e) \prec \tilde{\theta}(\cdot,e)$ and that $\tilde{\theta}(\cdot,e)$ is symmetric unimodal about a point. It can then be easily shown, using the orthogonal nature of matrix $\VEC A$, that  $\eta(\cdot,e) \prec \tilde{\eta}(\cdot,e)$ and that $\tilde{\eta}(\cdot,e)$ is symmetric unimodal about a point.
 We now use the result in Lemmas \ref{lemma:6.7} and \ref{lemma:new} to conclude that  $\mu \conv \eta(\cdot,e') \prec  \mu \conv \tilde{\eta}(\cdot,e')$ and that $\mu \conv \tilde{\eta}(\cdot,e')$ is symmetric unimodal about the same point as $\tilde{\eta}(\cdot,e')$. Thus, we have established that for all $e \in \mathcal{E}$, $\pi(\cdot, e) \prec \tilde{\pi}(\cdot,e)$. 
 
 To prove that $\tilde{\pi}(\cdot,e)$ is symmetric and unimodal about the same point it suffices to show that $\tilde\zeta(\cdot,e')$ are symmetric and unimodal about the same point. Since $\tilde\zeta(\cdot,e')$ is  convolution of $\tilde{\eta}(\cdot,e')$ and $\mu$, its symmetric unimodal nature follows from Lemma \ref{lemma:new}.  
\end{proof}
\begin{lemma}
Define $L(\theta) :=  \inf_{\VEC a \in \mathbb{R}^n} \mathds{E}[\norm{\VEC X_t- \VEC a}^2 |\Theta_t=\theta]$. $L(\cdot)$ satisfies Property $\RELATION^n$.
\end{lemma}
\begin{proof}
Let $\theta \RELATION^n \tilde\theta$ such that $\tilde\theta(\cdot,e)$ is symmetric unimodal about $\VEC b$ for all $e$.
For any $\VEC a \in \mathbb{R}^n$, the conditional expectation in the definition of $L(\theta)$ can be written as
\begin{align}
&\sum_{e \in \mathcal{E}}\int_{\VEC x \in \mathbb{R}^n}\norm{\VEC x-\VEC a}^2\theta(\VEC x,e)
\end{align}
Consider any $e$ with positive probability under $\theta$ (that is, $\int_{\VEC{x} \in \mathbb{R}^n} \theta(\VEC x,e) > 0$). For a constant $c>0$, consider the function $\nu_c(\VEC x) = c-\min\{c,\norm{\VEC x-\VEC a}^2\}$. 
Then,
\begin{align}
&\int_{\VEC x \in \mathbb{R}^n}\nu_c(\VEC x)\theta(\VEC x,e) \leq \int_{\VEC x \in \mathbb{R}^n}\nu_c^{\sigma}(\VEC x)\theta^{\sigma}(\VEC x,e) = \int_{\VEC x \in \mathbb{R}^n}(c-\min\{c,\norm{\VEC x}^2\})\theta^{\sigma}(\VEC x,e) \label{eq:hardy1}
\end{align}
where we used Lemma \ref{lemma:11} in \eqref{eq:hardy1}. Using the fact that $\theta(\cdot,e) \prec \tilde\theta(\cdot,e)$ and Lemma~\ref{lemma:10}, we have
\begin{align}
&\int_{\VEC x \in \mathbb{R}^n}(c-\min\{c,\norm{\VEC x}^2\})\theta^{\sigma}(\VEC x,e) \leq \int_{\VEC x \in \mathbb{R}^n}(c-\min\{c,\norm{\VEC x}^2\})\tilde\theta^{\sigma}(\VEC x,e) \notag \\
&= \int_{\VEC x \in \mathbb{R}^n}(c-\min\{c,\norm{\VEC x-\VEC b}^2\})\tilde\theta(\VEC x,e),
\end{align}
where $\VEC b$ is the point about which $\tilde\theta$ is symmetric unimodal.
Therefore, for any $\VEC a \in \mathbb{R}^n$,
\begin{align}
&\int_{\VEC x \in \mathbb{R}^n}(c-\min\{c,\norm{\VEC x-\VEC a}^2\})\theta(\VEC x,e) \leq \int_{\VEC x \in \mathbb{R}^n}(c-\min\{c,\norm{\VEC x-\VEC b}^2\})\tilde\theta(\VEC x,e) \notag \\
& \implies  \int_{\VEC x \in \mathbb{R}^n}(\min\{c,\norm{\VEC x-\VEC a}^2\})\theta(\VEC x,e) \geq \int_{\VEC x \in \mathbb{R}^n}(\min\{c,\norm{\VEC x-\VEC b}^2\})\tilde\theta(\VEC x,e)
\end{align}
As $c$ goes to infinity, the above inequality implies that
\begin{equation} \label{eq:err_inequality}
 \int_{\VEC x \in \mathbb{R}^n}(\norm{\VEC x-\VEC a}^2)\theta(\VEC x,e) \geq \int_{\VEC x \in \mathbb{R}^n}(\norm{\VEC x-\VEC b}^2)\tilde\theta(\VEC x,e)
 \end{equation}
 Summing up \eqref{eq:err_inequality} for all $e$ establishes that
 \begin{align}
 \sum_{e \in \mathcal{E}}\int_{\VEC x \in \mathbb{R}^n}\norm{\VEC x-\VEC a}^2\theta(\VEC x,e) \geq \sum_{e \in \mathcal{E}}\int_{\VEC x \in \mathbb{R}^n}\norm{\VEC x-\VEC b}^2\tilde\theta(\VEC x,e),
 \end{align}
Taking infimum over $a$ in the LHS of the above inequality proves the lemma. 
\end{proof}
Thus, both terms in \eqref{eq:claimeq1} satisfy Property $\RELATION^n$ and hence $V_t$ satisfies Property $\RELATION^n$.

\textbf{Step 2:} If $V_{t}$ satisfies Property $\RELATION$, we will show that $W_t$ satisfies Property $\RELATION$ too. \\
Consider two distributions $\pi$ and $\tilde\pi$ such that $\pi \RELATION \tilde \pi$ and $\tilde \pi$ is symmetric unimodal about $\VEC b$. Recall that \eqref{eq:mdpeq2} defined $W_t(\pi)$ as
\begin{align}
&W_{t}(\pi) =  \inf_{\hat\gamma}\mathds{E}[c\ind_{\{U_t=1\}}+V_{t}(\Theta_{t})|\Pi_t=\pi,\gamma_t = \hat\gamma] \label{eq:gclaimeq2} =: \inf_{\hat\gamma} \mathds{W}(\pi,\hat\gamma)
\end{align}
For any $\gamma$,  we will construct another prescription $\tilde\gamma$ such that $\mathds{W}(\tilde\pi,\tilde\gamma) \leq \mathds{W}(\pi,\gamma)$. This will imply that $W_t(\tilde\pi) \leq W_t(\pi)$, thus establishing the statement of step 2. We start with 
   \begin{align}
&\mathds{W}(\pi,\gamma) =  \mathds{E}[c\ind_{\{U_t=1\}}+V_{t}(\Theta_{t})|\Pi_t=\pi,\gamma_t = \gamma] \notag\\
&= c\sum_{e}\int_{\VEC x}\pi(\VEC x,e)\gamma(\VEC x,e) + \mathds{E}[V_{t}(Q^2_t(\pi,Y_t,\gamma))|\Pi_t=\pi,\gamma_t = \gamma]  \label{eq:gdpeq2aux1}
\end{align}
The second term in \eqref{eq:gdpeq2aux1} can be further written as
\begin{align}
&=\sum_{e'}\int_{\VEC x'}\pi(\VEC x',e')(1-\gamma(\VEC x',e')) \times V_t(\theta^{\gamma}) +\sum_{e}\int_{\VEC x}\pi(\VEC x,e)\gamma(\VEC x,e)V_t(\delta(\VEC x,e-1)) \label{eq:gdepeq2aux2}
\end{align}
where $\theta^{\gamma}$ is the distribution resulting from $\pi$ and $\gamma$ when $Y_t =\epsilon$ (see Lemma \ref{lemma:Gupdate}). Substituting \eqref{eq:gdepeq2aux2} in \eqref{eq:gdpeq2aux1} and using the fact that $V_t(\delta(\VEC x,e-1)) = K(e-1)$ gives 
\begin{align}
&c\sum_{e}\int_{\VEC x}\pi(\VEC x,e)\gamma(\VEC x,e) + \sum_{e}\int_{\VEC x}\pi(\VEC x,e)\gamma(\VEC x,e)K(e-1) \notag \\
&+ \sum_{e'}\int_{\VEC x'}\pi(\VEC x',e')(1-\gamma(\VEC x',e')) \times V_t(\theta^{\gamma})
\end{align}
We define $\lambda(e):= \int_{\VEC x}\pi(\VEC x,e)(1-\gamma(\VEC x,e))$.  We construct $\tilde\gamma$ as  follows. 
Define $r \geq 0$ to be the radius of an open ball centered at $\VEC b$ such that  $\int_{\norm{\VEC x-\VEC b}<r}\pi(\VEC x,e) = \lambda(e)$. Then, define
\begin{align}
\tilde\gamma(\VEC x,e) = \left \{ \begin{array}{ll}
               0 & \mbox{if $\norm{\VEC x-\VEC b} < r$} \\
                  1 & \mbox{otherwise} 
                \end{array}
               \right. \label{eq:Gthreshextra}
          \end{align}
Using the expressions for $\theta^{\gamma}$ and $\tilde\theta^{\tilde\gamma}$ from Lemma \ref{lemma:Gupdate} and the fact that $\pi \RELATION \tilde\pi$, it can be shown that $\theta^{\gamma} \RELATION \tilde\theta^{\tilde\gamma}$. This establishes the result of Step 2. 

                                  


\bibliographystyle{IEEEtran}
\bibliography{myref,collection}
\end{document}